# Probing Phonon dynamics and Electron-Phonon Coupling by High Harmonic Generation in Solids

Shi-Qi Hu[1,2], Hui Zhao[1,2], Xin-Bao Liu[1,2], Da-Qiang Chen[1,2], and Sheng Meng[1,2,3*]

[1]Beijing National Laboratory for Condensed Matter Physics and Institute of Physics, Chinese Academy of Sciences, Beijing, 100190, P. R. China

[2]School of Physical Sciences, University of Chinese Academy of Sciences, Beijing, 100049, P. R. China

[3]Songshan Lake Materials Laboratory, Dongguan, Guangdong 523808, P. R. China

*Author to whom correspondence shall be addressed: smeng@iphy.ac.cn

**Abstract:** Acting as a highly nonlinear response to the strong laser field, high harmonic generation (HHG) naturally contains the fingerprints of atomic and electronic properties of materials. Electronic properties of a solid such as band structure and topology can thus be probed, while the phonon dynamics during HHG are often neglected. Here we show that by exploiting the effects of phonon deformation on HHG, the intrinsic phonon information can be deciphered and direct probing of band- and mode-resolved electron-phonon couplings (EPC) of photoexcited materials is possible. Considering HHG spectroscopy can be vacuum free and unrestricted to electron occupation, this work suggests HHG is promising for all-optical characterization of EPC in solids, especially for gapped quantum states or materials under high pressure.

The high harmonic generation (HHG) in solids is a new frontier which has stimulated numerous advances in attosecond science and strong-field condensed-matter physics [1,2]. Originating from the nonlinear driving of electrons within and between electronic bands by strong-field light-matter interactions [3-5], the HHG spectra naturally contains significant fingerprints of intrinsic atomic and electronic properties of materials. Only recently, studying these strong-field phenomena in condensed matter systems has been realized, and a lot of excitement is emerging in the community about learning about the material's properties through this nonlinear non-perturbative laser-matter interaction, including the lattice structure [6,7], electronic topology [8-10], dynamic conductivity [11] and band structure [10,15,16].

Unfortunately, in previous studies, due to its slower timescale (lasting sub-picoseconds), the effect of atomic dynamics on solid HHG (usually on the timescale of sub-femtoseconds) [6,8-12] is often ignored, and thus the HHG in solids has hardly been applied to analyze the phonon dynamic effects on the attosecond electronic response, though the coherent excitation of phonons by infrared laser becomes available [17,18]. This gap stands in stark contrast to gas-phase molecules, where the role of atomic displacement (vibration) on HHG has been extensively studied. For example, pioneering studies have demonstrated the relationship between the nuclear autocorrelation function and HHG [19,20]; coupled electronic and nuclear dynamics have been probed by HHG spectroscopy [21]; as well as coherent phase and amplitude modulation of XUV emitted from vibrating molecules were demonstrated [22].

In reality, solid HHG, taking place in the bath of quasiparticles including laser-stimulated coherent phonons (CPs) which can significantly alter the material properties such as lattice structure and phase transition [23,24], is originated from electron motions in a varying lattice potential modulated by phonon deformations (such as bond compression/stretching dynamics). Very recently, the phonon dynamics is shown to significantly affect the nonlinear electron dynamics and thus time-dependent HHG, producing sidebands, plateau modulations and CEP dependence on the harmonic spectra [25-28]. These studies imply a great scientific/technological prospect on tracking the interplay of electronic and lattice dynamics. However, the rational analysis on the electronic mechanism behind the phonon-modulated HHG processes and thus the simultaneous probing of the electron-phonon coupling (EPC), one of the most important interactions in solids, is still lacking.

In this work, the temporally resolved electronic structure and phonon-deformation effects are revealed and analysed via detecting HHG spectra within a phonon period in a prototype transition-metal dichalcogenide MoS$_2$. A direct probing method on the band- and mode-resolved EPC matrix element $g_{ii}$ for unoccupied electronic states is established and validated by the state-of-the-art first-principles simulations. Our study reveals that the many-body interactions (i.e. the EPC strength) can be probed during attosecond dynamics in condensed matters; it also provides an alternative all-optical route for characterizing phonon dynamics and EPC complementary to angle-resolved photoemission spectroscopy (ARPES). Our results are helpful to explore EPC-related physics by HHG, such as microscopic mechanisms of superconductivity [29,30] and polaron formation [31,32], especially for gapped quantum states and materials under high pressure where ARPES are hardly applicable to measure such quantities.

Our study is performed based on the real-time time-dependent density-functional theory molecular dynamics (rt-TDDFT-MD) [33,34], which has successfully described the dynamics of light-matter interactions [35,36]. To study the HHG response, a linearly polarized laser pulse with a time-dependent electric field $\boldsymbol{F}(t) = \boldsymbol{F}_0 \cos(\omega t) \exp(-(t-t_0)^2/2\sigma^2)$ is adopted. The full width at half maximum (FWHM), $\sigma$, is about 50 fs for phonon pump laser and 20 fs for probe laser generating HHG. By adding the vector potential $\boldsymbol{A}(t) = -c \int \boldsymbol{F}(t')dt'$ in the Hamiltonian and evolving the electronic wavefunctions in real time, the time-dependent current is obtained as $\boldsymbol{j}(t) = \frac{1}{2i} \int dr \, [\psi^*(r,t)\nabla\psi(r,t) + c.c.]$. The HHG spectra are thus obtained by the Fourier transform of photocurrent, $\text{HHG}(\omega) = |\int \omega^2 \boldsymbol{j}(t) \exp(-i\omega t)dt|^2$.

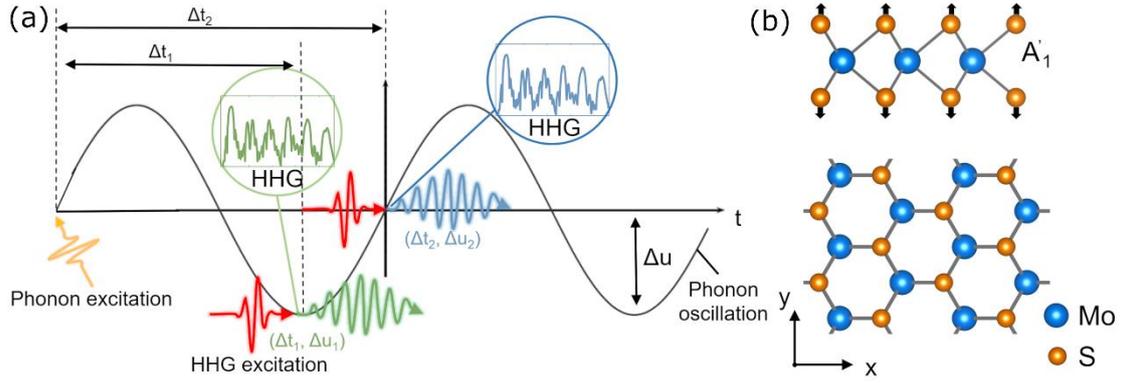

**Fig. 1. (a)** Schematics showing the pump-probe setup for coherent phonons (CPs) and high harmonic generation (HHG). The coherent phonon is excited and subsequently probed by the HHG with different phonon deformation at different pump-probe delay. **(b)** Schematics showing the atomic structure and $A_1'$ phonon mode in monolayer $MoS_2$.

The calculated HHG in monolayer $MoS_2$ is shown in Fig. 1(a), where a ~ 1.7 eV 50-fs pump pulse with peak field strength of 0.036 V/Å, polarization along *y* direction is used to excite CPs in $MoS_2$, then subsequently an intense IR laser pulse (~0.3 eV, 20 fs) is applied to probe the phonon dynamics by generating HHG. The polarization of probe laser electric field is along the *x* direction (**Γ-K** direction). To compare with the recent experiments [17,37], the photon energy of pump and probe lasers identical to experimental ones are used in our investigation.

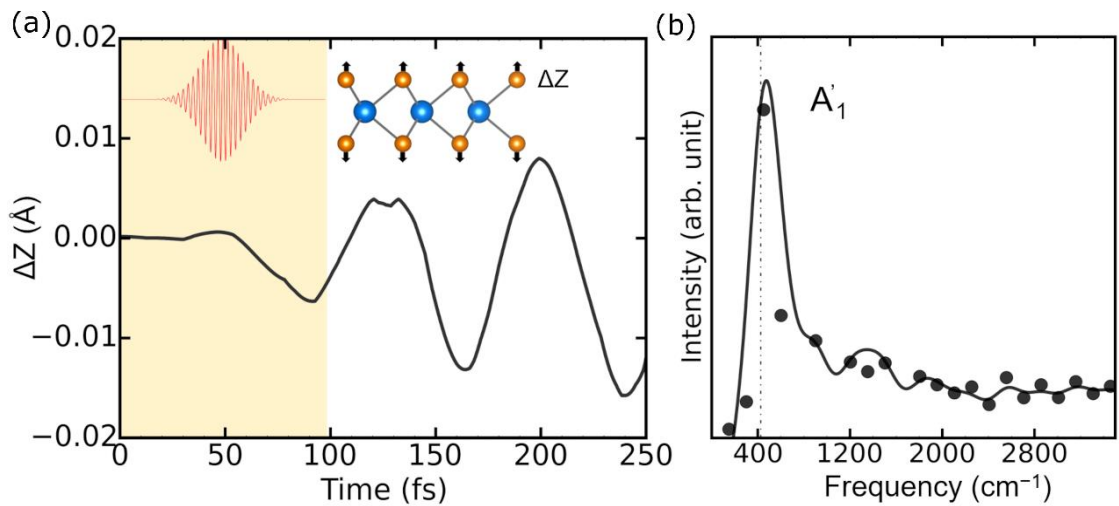

**Fig. 2. (a)** The displacements of S atom along the $A_1'$ phonon eigenmodes as a function of time after photoexcitation. The red line shows the configuration of pump laser. **(b)** The Fourier analysis of photoexcited atomic oscillations.

Figure 2 shows the laser-induced CPs of monolayer MoS$_2$. Consistent with the experimental observation [17], the laser pulse mainly produces the A$_1^{'}$ mode with the S atom displaced along the $z$ direction (Fig. 1(b) and inset of Fig. 2(a)), which has an oscillation period ~78 fs, corresponding to the frequency ~420 cm$^{-1}$. The resultant maximum atomic displacement is ~0.01 Å.

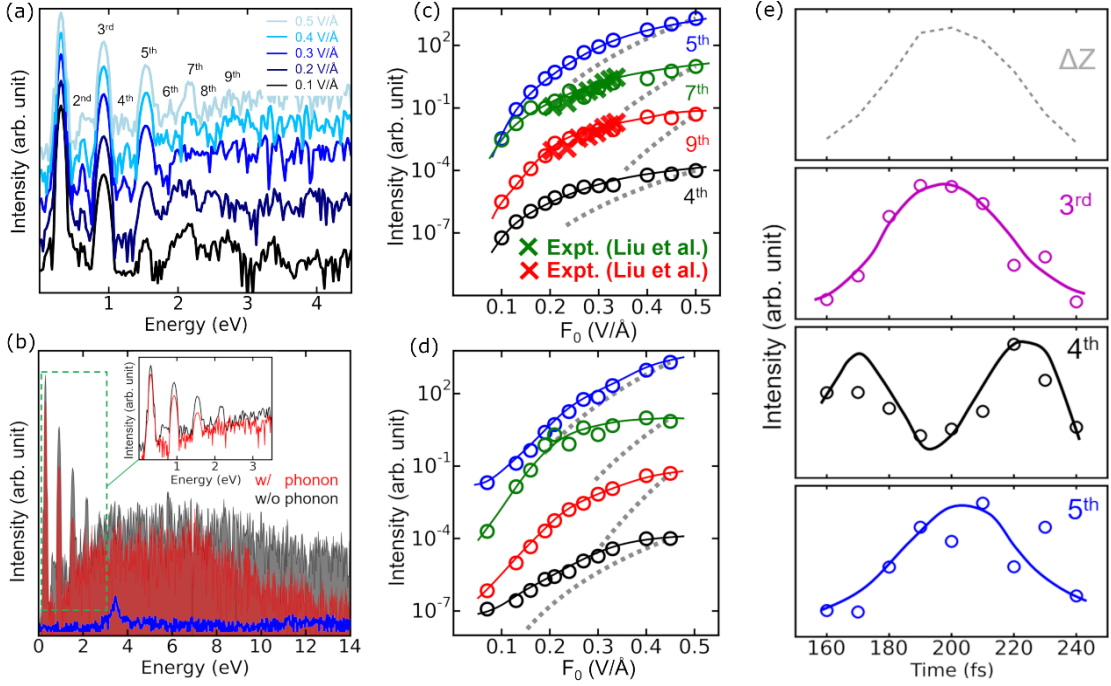

**Fig. 3. (a) The HHG spectra without CPs under different laser field. (b) The HHG spectra without (black) and with (red) A$_1^{'}$ CPs. The blue line shows the signal of phonon pump laser. The HHG intensity as a function of laser field without (c) and with (d) A$_1^{'}$ CPs. The dashed grey lines denote the perturbative nonlinear response $I_n \sim F_0^{2n}$. The black crosses represent the experimental data in ref. [37]. (e) Normalized HHG yield as a function of pump-probe delay time. The solid lines are guide to eyes.**

We then investigate the HHG process with the CPs. With no phonon vibrations (fixed lattice), Fig. 3(a) and 3(c) display the HHG spectra and intensity as a function of laser electric field intensity $F_0$. Due to the space inversion symmetry breaking, both the odd and even harmonics up to 9$^{th}$ order are obtained. As the field increases, more harmonic peaks appear in the HHG spectra, and the field strength dependence of $n$th order harmonic intensity $I_n$ shows an obvious deviation from the power law behavior $I_n \sim F_0^{2n}$ (the grey dashed lines), indicating a non-perturbative character of the HHG process. We also show the HHG data measured in the experiment [37] using a similar photon energy (~ 0.3 eV) and field strength, where a good agreement between the calculated and experimental HHG is found.

In contrast, with CPs pumped, the spectra become noisy and less frequency-comb like for both odd and even harmonic peaks and the intensity of HHG is decreased, as shown in Fig. 3(c). This feature is similar to the previous results of HHG in monolayer boron nitride [25], indicating that the temporal dynamics of electrons is no longer periodic due to the phonon deformation effects (i.e., the phonon-induced lattice displacement). Considering the weak intensity, different photon energy and polarization of the phonon pump laser, the HHG yielded here comes from the probe laser only. As shown in Fig .3 (c), the HHG yielded by the probe laser is much higher than the signal of the phonon pump laser (the blue line), indicating the pump laser has almost no disturbance on the HHG process.

On the other hand, considering the long period (~80 fs) and lifetime of coherent phonons (usually on the order of 10 picoseconds) compared to electron dynamics driven by laser (~20 fs), the interference of CPs dynamics to HHG process can be modulated by the time delay between the pump-probe time delay (Fig. 1(a)). With the excitation of CPs, we artificially change the delay time of the probe laser from ~160 fs to ~240 fs (which is much longer than the FWHM of phonon pump laser (~50 fs), indicating the HHG process is well separated from the pump laser field). The HHG yield of each order oscillates periodically with the CPs (Fig. 3(e)). Interestingly, the oscillation shows differences between odd and even harmonics--the former has an identical period with CPs while the later has the double frequency.

These results indicate that phonon deformation effects alter the ponderomotive energy of HHG process by modulating the electronic structure and effective electron mass. As shown in Fig. 1(a), the phonon-deformation effects on the HHG can be further understood from the electron structure mechanism as follows: with the specific phonon mode excited, the $j$th atom has the time-dependent displacement $\Delta \boldsymbol{u}_j(t)$. The $\Delta \boldsymbol{u}(t)$ (index $j$ is omitted hereafter) tunes the interatomic potential and thus the $i$th band level $\varepsilon_i$ through EPC. For example, at $t_1$, the atom is located at the equilibrium position with the displacement $\Delta \boldsymbol{u}(t_1) \sim 0$, there is no deformation effect and the band has the equilibrium value $\varepsilon_i(t_1)$. At a later time $t_2$, the atom moves to the maximum $\Delta \boldsymbol{u}(t_2)$, corresponding to a modulated band structure $\varepsilon_i(t_2)$ with phonon deformation. Since the HHG depends on the band structure, the strong-field-driven electron wavepacket in the modulated potential $\varepsilon_i(t_2)$ shows a different dynamic process with that in $\varepsilon_i(t_1)$, which gives the different currents and HHG spectra between time $t_1$ and $t_2$.

Therefore, besides paving the way to probe phonon dynamics and ultrafast lattice changes with sub-cycle temporal resolution [25], the phonon modulated HHG spectra also provide a potential approach for an all-optical detection on the fundamental EPC. According to the definition of EPC matrix element $g_{ii}$, i.e., $g_{ii} \sim \langle\varphi_i |\partial V/\partial \boldsymbol{u}|\varphi_i\rangle$, where $V$ is the deformation potential of electrons and $\boldsymbol{u}$ is atom displacement of a given phonon mode. The $g_{ii}$ can be calculated as the first derivative of the $i$th band energy $\varepsilon_i$ with respect to atomic displacement, i.e., $g_{ii} = \partial \varepsilon_i/\partial \boldsymbol{u} = \Delta \varepsilon_i/\Delta \boldsymbol{u}$ [38,39]. The equation indicates for a certain phonon amplitude (atomic displacement) $\Delta u$, the EPC scales with the band structure change $\Delta \varepsilon_i = \varepsilon_i(t_1) - \varepsilon_i(t_2)$. Therefore, EPC probing relies on the measurement on $\Delta \varepsilon_i$.

As proven in the previous study [37], under our laser condition, the HHG in MoS$_2$ is mainly contributed by the intraband current mechanism within the semiclassical scattering theory [12,40]:

$$\boldsymbol{j}(t) \propto v(t) = \frac{1}{\hbar}\frac{\partial \varepsilon(\boldsymbol{k})}{\partial \boldsymbol{k}} - \frac{\partial \boldsymbol{k}(t)}{\partial t} \times \boldsymbol{\Omega}(\boldsymbol{k}), \qquad (1)$$

where $\hbar \frac{\partial \boldsymbol{k}(t)}{\partial t} = -e\boldsymbol{F}$. The first and second term reflects nonlinear currents from band dispersion as well as the Berry curvature (Fig. 4(a) and (d)), which contribute the odd and even harmonics, respectively. The intraband mechanism is further confirmed by the time-frequency HHG spectrograms, where HHG is generated at the peak of the electric field, with no chirp (as shown in Fig. S1 in supplementary materials) [4,11,41,42].

For probing $\Delta\varepsilon_i$, the band structure can be retrieved from the odd HHG in MoS$_2$, as shown in Fig. 4(a): Under the strong field, the electron wavepacket is driven to oscillate in the energy bands and scattered due to the nonlinearity of band dispersion (the black circle in Fig. 4(a)), which is described by a series of spatial harmonics expansion $\varepsilon(\boldsymbol{k}) = \sum_{m=0}^{m_{max}} \beta_{i,m} \cos(m\boldsymbol{k}a)$. Here $a$ is the lattice constant, $m$ is the index of lattice site and $\beta_{i,m}$ is the Fourier coefficients of band dispersion. Then the EPC matrix element gives: $g(\boldsymbol{k}) \propto \sum_{m=0}^{m_{max}} \beta_{i,m}(t_1) \cos(m\boldsymbol{k}a) - \sum_{m=0}^{m_{max}} \beta_{i,m}(t_2) \cos(m\boldsymbol{k}a)$, which depends on $\beta_{i,m}$.

Since the dependence of the $n$th order harmonic intensity $I_n$ on the field strength $\boldsymbol{F_0}$ is proportional to [12,16,40]:

$$I_{n,i} \propto (n\omega)^2 \left|\sum_{m=1}^{m_{max}} m\beta_{i,m} J_n\left(\frac{eF_0 a}{\hbar\omega}\right)\right|^2, \qquad (2)$$

for a laser field with a frequency $\omega$. Where $J_n$ is the Bessel function of the first kind of order $n$. The band dispersion coefficients $\beta_{i,m}$ can be obtained by fitting the field dependence of HHG intensity [12,16]. This scenario is proposed to explain well the behaviour and spatial properties of HHG in solid SiO$_2$[12], which has been shown to be capable to reproduce the weak carrier envelope phase sensitivity and chirp-free emission [11,12]. The related technique based on HHG spectra enabled the complete, all-optical reconstruction of the band dispersion profile by mid-infrared laser produced HHG in insulator ZnO and ZnSe [15,16].

The measurements were in remarkable agreement with theoretical predictions. Very recently, the all-optical reconstruction by HHG is successfully used to retrieve the dispersion of multiple bands and Berry curvature of semimetal in large Brillouin zone space [10], demonstrating this scenario is well-established and robust.

By controlling pump-probe time delays $\Delta t = t_1$ and $t_2$, and fitting the HHG intensity as a function of field strength with Eq. (2), the different band structures without/with the phonon deformation $\varepsilon_i(t_1)$, $\varepsilon_i(t_2)$, and thus $\Delta\varepsilon_i$ can be measured in our investigation. The resolution of the probing method is determined by the ratio between phonon period and laser FWHM $\sigma$. The probing method is valid when the HHG can be detected within less than the half phonon period, i.e., $\sigma < 1/\hbar\omega_{ph}$. In our study the $\sigma$ ~20 fs and the phonon period $1/\hbar\omega_{ph}$ is ~80 fs, which is around 4 times longer than the laser duration and thus the probing resolution is guaranteed.

Based on the above strategy, the 1$^{st}$ CB along **Γ**-**K** without/with $A_1'$ phonon deformation in MoS$_2$ is reconstructed in Fig. 4(b). To compare the band structure with and without the phonon deformation, we set the 1$^{st}$ CB maximum without the phonon deformation as 0 eV and use it to benchmark the energy of bands. All of the bands reconstructed from the HHG data with/without phonon deformations above agree well with the bands obtained from DFT. The bandwidth modulation due to the phonon deformation is also well reproduced, demonstrating the effectiveness of the present approach.

Combining with the time-resolved X-ray/electron diffraction [43,44] technology, the corresponding atomic displacement $\Delta u$ can be directly measured from the diffraction intensity (seeing Fig. S2 in supplementary materials [41,45]) $I_{hkl}(\boldsymbol{x},\boldsymbol{y},\boldsymbol{z}) \propto |F_{hkl}(\boldsymbol{x},\boldsymbol{y},\boldsymbol{z})|^2$, where $F_{hkl} = \sum_j f_j \exp[2\pi i(\boldsymbol{h}x_j + \boldsymbol{k}y_j + \boldsymbol{l}z_j)]$

is the structure factor and $x_j$, $y_j$, $z_j$ are the positional coordinates of the $j$th atom, $f_j$ is the scattering factor of the $j$th atom and can be found in ref. [46].

We then further estimate the distribution of mode-resolved EPC matrix element $g_{ii}$ of the 1st CB along **Γ**-**K**. As shown in Fig. 4 (c), the configuration of $g_{ii}$ probed by the HHG is in good agreement with that calculated by density functional perturbation theory (DFPT), demonstrating the EPC measurement by time-dependent HHG is effective.

Considering the HHG process is mainly contributed by the low conduction band or high valence band near the Fermi level, realizing the band and phonon mode selectivity of EPC probing by HHG can be further improved, which may be achieved by switching the polarization or wavelength of driving laser.

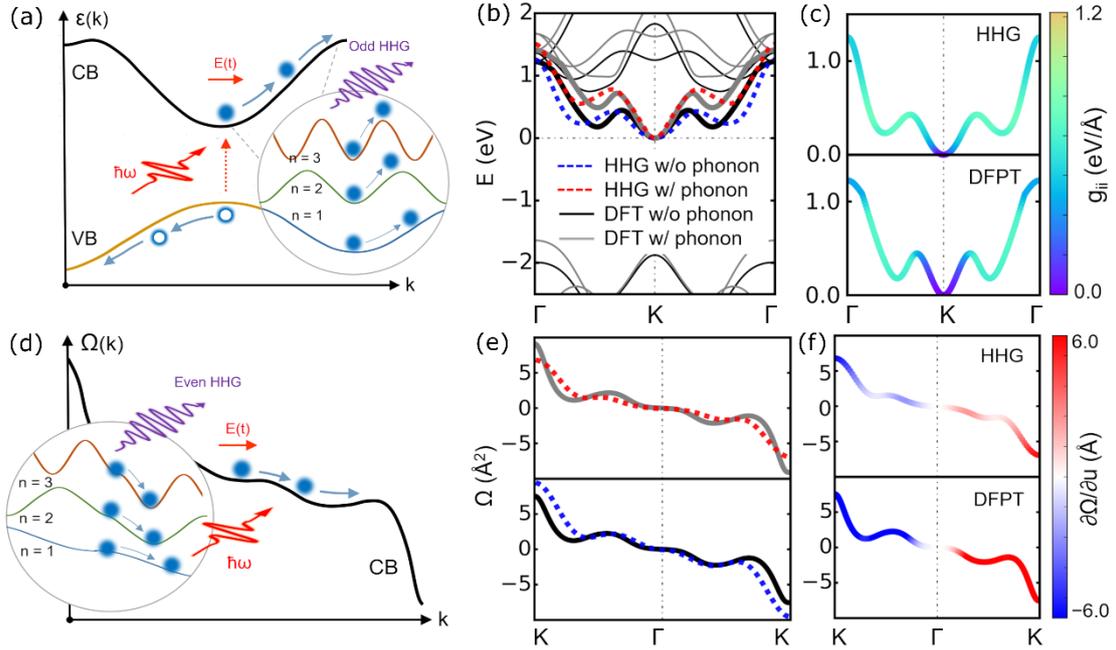

**Fig. 4. Semiclassical picture of the field-driven electron dynamics in bands (a) and Berry curvature (d). The HHG can be generated by the modulation of the intraband current due to the nonlinearity of electron band and Berry curvature (as shown in the black cycles). The band structure (b) and Berry curvature (e) along Γ-K direction without and with the phonon deformation. The black and grey lines are the DFT result while the red and blue dash lines are retrieved from the**

**HHG. (c) The distribution of band- and mode-resolved EPC matrix element $g_{ii}$ and (d) phonon response of Berry curvature $\partial\Omega/\partial u$ along $\Gamma$-$K$ direction.**

Similar to the previous methodology [10, 47], the Berry curvature $\boldsymbol{\Omega}$ of the 1st CB along $\boldsymbol{\Gamma}$-$\boldsymbol{K}$ without/with $A_1$' phonon deformation in MoS$_2$ can also be probed by even order harmonics in Fig. 4(e). The absolute magnitude has been determined using the formula $|\beta_{i,m}|na \approx \boldsymbol{F}_0|\gamma_{i,m}|$, as used in ref. [47], where $\gamma_{i,m}$ is the Fourier coefficients of Berry curvature: $\Omega(\boldsymbol{k}) = \sum_{m=0}^{m_{max}} \gamma_{i,m}\sin(mka)$. The response of Berry curvature due to the $A_1$' phonon deformation, i.e., $\partial\boldsymbol{\Omega}/\partial\boldsymbol{u}$, is correspondingly retrieved, with $\boldsymbol{\Omega}$ and $\partial\boldsymbol{\Omega}/\partial\boldsymbol{u}$ show good agreements between the HHG probed and DFPT results.

Since the EPC probing by HHG is not restricted to occupied electronic bands and can be vacuum free, we propose that this approach is particularly suitable for EPC probing of materials with wide bandgaps such as insulating oxides and materials at extreme conditions such as the high-pressure and room-temperature superconductors including H$_3$S and LaH$_{10}$[48,49]. Here we explore this scenario in Im-3m H$_3$S at ~200 GPa, which has superconductivity at 200 K [48]. As shown in Fig. S3 of supplementary materials [41], the EPC near Fermi level is well retrieved by the phonon-modulated HHG. Other examples could include light-induced potential superconductors such as YBa$_2$Cu$_3$O$_{6.5}$ and K$_3$C$_{60}$ [50,51], where the superconductivity mechanism is still unclear due to the lacking of effective EPC detection methods.

In conclusion, our study shows that CPs would effectively modulate the ultrafast HHG processes. Based on the phonon-deformation effects on HHG, we show that by applying the pump-probe setting, the phonon dynamics, mode-resolved EPC and Berry curvature response can be effectively probed with an all-optical HHG approach, providing a promising alternative route for EPC characterization complementary to ARPES. This finding paves the way to the real-time measurement of nonequilibrium EPC with a subcycle temporal resolution, suggesting a powerful tool to investigate the ultrafast electron-nuclear coupled dynamics.

**Acknowledgements**

We acknowledge financial support from MOST (grant 2021YFA1400200), NSFC (grants 12025407 and 11934003), and CAS (XDB330301).

**Author contributions**

S. M. conceived and supervised the project. S.-Q. H. performed the theoretical modelling and TDDFT calculations. D.-Q. C contributed to the software code. S.-Q. H., H. Z, X.-B. L., and S. M. analyzed the data. The manuscript is written by S.-Q. H. and S. M.

**Data availability**

The data that support the findings of this study are available from the corresponding author upon reasonable request.

**Competing interests**

The authors declare no competing financial or non-financial interests.